  \documentclass[jcph]{apjrnl}

  {\theoremstyle{plain}%

  {\theoremstyle{remark}

  }
  {\theoremstyle{definition}

  }
\begin{document}

  %% Author, fill in article title here
  \title{\bf Energy minimization using Sobolev gradients: application
  to phase separation and ordering}
  \thanks{Acknowledgements.  We are grateful to NSERC of Canada and
  Los Alamos National Laboratory for support. We  thank SHARCNET
  for use of  computing resources.}

  %% Fill in author list here
  \author{%
  S. Sial\ref{first},
  J. Neuberger\ref{second},
  T. Lookman\ref{first},\ref{third} and A. Saxena\ref{third}\\%
  %J. Neuberger\ref{second}, and A. Saxena\ref{third}\\%
  \additem[first]{University of Western Ontario, Department of Applied
  Mathematics, London, Ontario, Canada N6A 5B7},
  \additem[second]{University of North Texas, Department of Mathematics,
  Denton, Texas 76203-1430}
  \additem[third]{ Theoretical Division, Los Alamos National Laboratory, MS
  B262, New Mexico 87545}
  \\E-mail: ss@cmrg.apmaths.uwo.ca, jwn@unt.edu, txl@viking.lanl.gov,
  abs@viking.lanl.gov }

  %\date{24 June 2002} %please do not use \today, use actual date of version

  \maketitle

  \begin{abstract}
  A common problem in physics and engineering is the calculation of
  the minima of energy functionals.
  The theory of Sobolev gradients provides an efficient method
  for seeking the critical points of such a functional. We apply the method
  to functionals describing coarse-grained Ginzburg-Landau models commonly
  used in pattern formation and ordering processes.
  \end{abstract}


  %%%% Authors begin text of article here %%%
  %%%% Authors begin text of article here %%%
  %%%% Authors begin text of article here %%%

  \section{\bf Introduction }

  Many problems in mathematical physics are formulated in terms of finding
  critical points of energy functionals.  The recent theory of Sobolev
  gradients \cite{SV} provides a unified point of view on such problems,
  both in function spaces and in finite dimensional approximations to
  such problems. The aim of this work is to demonstrate the use and
  efficiency of Sobolev gradient techniques in minimising energy
  functionals associated with Ginzburg-Landau models for studying phase
  transitions in alloys and complex fluids. These equations are
  prototypical for studying pattern formation or ordering, such as
  nucleation and spinodal decomposition, that are accompanied by instabilities.
  We illustrate our work with models A and B in the Halperin-Hohenberg
  taxonomy, in which the coarse-grained field or the order parameter ($OP$)
  is either not conserved (model A) or  conserved (model B) \cite{Hohenberg}.

  A gradient of a functional gives the direction of greatest change per
  unit change in the argument of the functional.  Often
  overlooked is that the direction of a gradient strongly depends on
  how the {\it size} of arguments of a functional are measured.
  Functionals of interest in physics, particularly energy functionals,
  commonly include derivatives of the arguments.  Such arguments have to
  be considered large if some of its derivatives are large.  Theoretical
  considerations of such functionals must take this into
  account but is often overlooked in numerical
  approximations.   The theory of
  Sobolev gradients \cite{SV} is an organized account of how to choose a
  metric for a finite dimensional problem that matches that {\it
  required} for the corresponding theoretical problem.  It is found
  that a proper matching leads to gradients (Sobolev gradients)
  which are considerably smoother than those normally used
  \cite{Richardson}.  The result is that the approach to a minimum
  energy configuration becomes much more efficient.   In fact, the
  improvement in performance using Sobolev gradients becomes infinite as
  mesh size  goes to zero.  This paper illustrates this phenomenon in
  some typical problems of interest in phase separation and pattern
  formation. The layout of the paper is as follows:
  An introduction to Sobolev gradients in Section 2 is followed by
  a description in Section 3 of Ginzburg-Landau models and how Sobolev gradient
  techniques may be employed. In Section 4 we compare the results for
  minimization using ordinary gradients (functional derivatives) and
  in an appropriate Sobolev space in 1, 2, and 3 dimensions, with
  different grid spacings and with different boundary conditions.
  If we consider steepest descent as being a time evolution from a
  higher energy state to a lower energy state, then a theoretical bound
  on how large our time step can be is given by the
  Courant-Freiderichs-Lewy (CFL) condition \cite{CFL}.  Beyond this limit,
  the numerical scheme for steepest descent will magnify errors in each
  step. This implies that for the traditional steepest descent schemes
  the step size will have to be decreased as grid spacing becomes finer,
  the dimension of the problem is increased or the order of the
  derivatives in the problem increases.  The
  Sobolev gradient technique avoids these problems \cite{Richardson}.
  When we use ordinary gradients we label our results "$L_2$" runs since
  ordinary gradients are closely related to attempts at defining a
  gradient in $L_2(\Omega)$, the space of square integrable functions in
  a region $\Omega$.
  In Section 5 we consider model A$^\prime$, which is model A with a
  constraint, namely, the average value of the $OP$ is conserved. This
  model has a different Sobolev gradient than model A and is an alternative
  to the Cahn-Hilliard equation (model B) when dynamics is not of interest.
  We compare results for minimization in $L_2$ using the Cahn-Hilliard approach
  to model A$^\prime$ minimization in the appropriate Sobolev space
  in 1, 2, and 3
  dimensions with different grid spacings and with different boundary
  conditions.
  In Section 6 we extend model A$^\prime$ to models of surfactant systems
  which have higher order derivatives.
  For the models we have studied, the Sobolev gradient technique
  becomes increasingly atttractive as grid spacing is refined, dimension
  is increased, or the order of the derivatives in the problem becomes
  higher.

  \section{\bf The Sobolev Gradient approach }

  Sobolev gradients essentially provide an organized numerical procedure
  of determining preconditioners.
  An  energy functional can be generically written as:
  $$
   J(u) = \int_{\Omega} F(Du),  \ \ \ \ \ \ \ \ \ \ (1)
  $$
  where $\Omega$ is a domain in  Euclidean space, $u$ is a member of an
  appropriate function space and $Du$ is a list of (length $n$, say)
  consisting of $u$ and all partial derivatives of $u$ which are relevant to
  the problem at hand.  $F$ is a function on an appropriate Euclidean space.
  For example, consider $\Omega$ to be a rectangular region
  in $R^2$,  $F$ is a function on $R^3$ so that
  $$
  F(w,r,s) = \frac{ (r^2+s^2) }{ 2 } + \frac{ w^4 }{ 4 } - \frac{ w^2 }{ 2 }
  $$
  for all numbers $w,r,s$ and $D$ is the transformation
  $
  Du = (u,u_x,u_y)
  $
  for all $u$ on
  $\Omega$ with well-defined partial derivatives.
  Equation (1) then takes the form
  $$
  J(u) = \int_{\Omega} F(Du) =
  \int \left [\frac{ (u_x^2 + u_y^2) }{ 2 } + \frac{ u^4 }{ 4 }
  - \frac{ u^2 }{ 2 } \right ] dr,
  $$
  which is one of the functionals we will deal with in this paper.
  Returning to our general considerations of (1), we perform
  a first variation
  $$
  J^{\prime}(u)h = \int_{\Omega} F^{\prime}(Du)Dh.
  $$
   At this point we depart from custom and do {\em not}
  integrate by parts to obtain the Euler-Lagrange equations. Instead, we write
  $$
   J^{\prime}(u)h = \int F^{\prime}(Du) Dh = \langle Dh,(\nabla F)(Du)
  \rangle_{L_2(\Omega)^n}. \ \ \ \ \ \ \ \ \ \ \ \ (2)
  $$
  We note that
  $$
  \langle Dh,Dg \rangle_{L_2(\Omega)^3} = \int_{\Omega} (hg + h_x g_x + h_y g_y),
  $$
  the inner product in the Sobolev space $H^{1,2}(\Omega)$
  \cite{SV} \cite{Adams}.

  By $L_2( \Omega )$ we mean the Hilbert space of real functions on
  $\Omega$ in which

  $$
  ||f||^2_{ L_2( \Omega ) } ~=~
  \int_{\Omega} f^2
  $$

  \noindent  By $H^{1,2} ( \Omega )$ we mean the subspace of
  $L_2( \Omega )$ consisting of all $f$ so that the norm

  $$
  ||f||^2_{ H^{1,2}( \Omega ) } ~=~
  \int_{\Omega} ( f^2 ~+~ f_x^2 ~+~ f_y^2 )
  $$

  \noindent is defined.

      We introduce a transformation $P$ which is
  essential to our presentation.  Take $P$ to be the orthogonal projection of
  $L_2(\Omega)^n$ onto the subspace of all elements of the form $Du$.
  Such a transformation can be dealt with in a concrete way computationally.
      From (2),
  $$
  J^{\prime}(u) h = \langle Dh,(\nabla F)(Du) \rangle_{L_2(\Omega)^n}
                  = \langle P Dh, (\nabla F)(Du) \rangle_{L_2(\Omega)^n}
                  = \langle Dh, P (\nabla F)(Du) \rangle_{L_2(\Omega)^n}.
  $$
     These are legitimate steps since $P Dh = Dh$ and orthogonal projections may
  be
  passed form one side of an inner product to the other.
      We need one more inner product:
  $$
  \langle g,h \rangle_S = \langle Dg,Dh \rangle_{L_2(\Omega)^n}.
  $$
       In terms of this inner product,
  $$
  J^{\prime}(u) h = \langle Dh,P (\nabla F)(Du) \rangle_{L_2(\Omega)^n}
                  = \langle h, (\nabla_S J)(u) \rangle_S,
  $$
  where $(\nabla_S J)(u)$ is defined as the first element in the list
  $$
  P (\nabla F)(Du).
  $$

  \noindent The function $(\nabla_S J)(u)$ is called the Sobolev gradient of $J$
  at the
  element $u$.
  To make the above calculations useful the projection $P$ must be presented in
  a suitable form and the relevant details are given later.
  In a number of previous applications of the methods (e.g., transonic flow
  \cite{SV},
  Ginzburg-Landau functionals for superconductivity \cite{SV}) it has been
  known that Sobolev gradients
  give vastly superior results to those obtained with ordinary gradients.
  In what follows,
  slight variations of the above will be used,
  these variations take into
  account a variety of boundary and other external conditions.

  % The main reference for Sobolev gradients is \cite{SV}.

  \section{\bf Application to Ginzburg-Landau models }

  Models A and B are defined by the equations
  $\frac{\partial u}{\partial t} = -\frac{\delta J}{\delta u}$ and
  $\frac{\partial u}{\partial t} = \nabla^{2}\frac{\delta J}{\delta u}$
  respectively, where J is a free energy functional. The static and
  dynamical properties of these models
  have been extensively studied, primarily in numerical work
  related to coarsening and growth of domains \cite{Hohenberg, Rogers}.
  The functional $J(u)$ usually has
  a polynomial form that depends on the nature of
  the phase transition as the coefficient of the quadratic term
  changes sign (as a function of temperature, pressure or some other
 thermodynamic variable). The widely used form with terms in $u^2$ and $u^4$ is
  associated with a second order or continuous transition, where there is no
  jump discontinuity such as latent heat.

  We seek to minimize the model A free energy functional

  $$
  J(u) ~=~ \int \left [ \frac{1}{4} u^4 ~-~ \frac{1}{2} u^2 ~+~ \frac{\kappa}{2}
 |\nabla u|^2 \right ] dr
  $$

  \noindent over some volume subject to certain boundary conditions.  The
  coefficient $\kappa$ determines the energy penalty for interfaces.

  In one dimension the problem can be reformulated as minimization of

  $$
  J(u_0 , u_1) ~=~ \int \left [ \frac{1}{4} u_0^4 ~-~ \frac{1}{2} u_0^2 ~+~
  \frac{\kappa}{2} u_1^2 \right ] dx
  $$

  \noindent subject to the constraint that the $L_2( \Omega )$
  functions $u_0$ and $u_1$ are of the form

  $$
  ( u_0 , u_1 ) ~=~ ( f , f_x )
  $$

  \noindent for some $H^{(1,2)}( \Omega )$ function $f$.  We seek a projection
  operator that maps $(u_0,u_1)$ in $L_2( \Omega ) \times L_2( \Omega )$ to the
  closest point in the subspace
  consisting of points of the form $(f,f_x)$.  This is given by minimizing
  the integral

  $$
  I ~=~ \int \left [( f ~-~ u_0 )^2 ~+~ ( f_x ~-~ u_1 )^2 \right ] dx
  $$

  \noindent over the interval subject to specified constraints.  Minimizing $I$
  gives the condition

  $$
  (1 ~-~ \partial_x^2 ) f ~=~ u_{0} ~-~ \partial_x u_1.
  $$

  A steepest descent scheme in $L_2 ( \Omega ) $ would be of the form

  $$
  u ~\rightarrow~ u ~-~ \lambda \nabla J(u),
  $$

  \noindent where $\lambda$ is some scalar and $\nabla J(u)$ is the variation
  of $J$ with respect to $u$ subject to boundary conditions.
  We instead perform a steepest descent in the space where the gradient
  is given by the projection we already found:

  $$
   \nabla J( u_0 , u_1 ) ~=~
   (1 ~-~ \partial_x^2 )^{-1}
   \left( \frac{\partial J(u_0,u_1)}{\partial u_0}  ~-~
      \partial_x \frac{\partial J(u_0,u_1)}{  \partial u_1}  \right).
  $$

  This is equivalent to changing the norm of candidate functions from

  $$
  ||f||^2 ~=~ \int f^2 dx
  $$

  \noindent to

  $$
  ||f||^2 ~=~ \int f^2 dx ~+~ \int f_x^2 dx
  $$

  \noindent  again subject to appropriate constraints such as
  boundary conditions.

  \section{\bf Results for model A }

  In this section we report results for model A in one dimension with
  periodic and Dirichlet boundary conditions.  The coefficient $\kappa$
  was set to 1.0 for all the numerical trials.  For periodic boundary
  conditions, systems of $M$ nodes
  with spacing $h$ were set up with random initial values for the order
  parameter $u$ such that the avaerage value $\langle u \rangle =0.05$
  at $t=0$.  The final minimum energy configuration
  should have $u=1.0$ everywhere.  The number of iterations, the largest
  step $\lambda$ that could be used, and the CPU time to obtain $u > 0.99$
  everywhere in the system are noted in the Tables.  The next three
  entries in the Tables are the number of iterations, step, and CPU time
  required when using the Sobolev gradient technique.
  For Dirichlet boundary conditions the order parameter $u$ was set to
  $0.01$ everywhere except at the ends where $u$ was fixed at zero.  The
  program was terminated when the magnitude of the  $L_2$ gradient was
  less than $10^{-5}$ everywhere in the system.

  When minimizing in $L_2( \Omega )$ we note that the largest step size
  that can be used for each minimization step decreases as the grid spacing is
  halved, as is implied by the CFL condition.  However, steepest descent
  using the Sobolev gradient does not suffer from this limitation.
  At each minimization step we first estimate the usual $L_2$ gradient, using
  finite differences to estimate derivatives.  Thus, for model A we
  estimate $\nabla F = u^3 - u - \nabla^2 u$.
  Using the Sobolev gradient the energy is minimized by a step
  $u \rightarrow u - \lambda * \nabla_S F$, where $\nabla_S F$ is the Sobolev
  gradient we want to use.
  Thus, at each minimization step we need to find the Sobolev gradient, given
  the usual $L_2$ gradient.
  This Sobolev gradient satisfies the linear equation
  $( 1 - \nabla^2 ) \nabla_S F = \nabla F$.
  This is solved iteratively.  The first time we need to calculate
  the Sobolev gradient we do not have a good initial guess, however,
  in subsequent iterations the Sobolev gradient serves as a good initial guess.
  The Sobolev gradients vary
  smoothly as the minimization progresses and so an iterative procedure
  is less costly computationally than using a direct solver each time.
  Since the operator $( 1 - \nabla^2)$ is symmetric, positive definite, we use a
  conjugate gradient solver.  Steepest descent and Jacobi iteration
  result in longer run times.

  Results are reported for a single Dec Alpha EV68 CPU.  The difference in
  codes for the $L_2$ minimization and the Sobolev space minimization is that
  in the case of the Sobolev space minimization a call to a solver that
  estimates the Sobolev gradient, given the $L_2$ gradient, is made and then the
  Sobolev gradient is used instead of the $L_2$ gradient.

 \newpage
  {\bf  One dimensional model A }

  \bigskip

  {\it  Periodic boundary conditions (BCs):}

  \smallskip
  \begin{tabular}{|c|c|c|c|c|c|c|c|}\hline
  \hline
  Nodes $M$ & spacing $h$ & iterations ($L_2$) & step  $\lambda$ ($L_2$) &
  CPUs ( $L_2$ ) & iterations & step $\lambda$ & CPUs       \cr
  \hline
  \hline
  $2^{10}$    & 1.0         & 18         & 0.32           & 0.00391
  & 10       & 0.6           & 0.0195     \\
  $2^{11}$    & 0.5         & 48         & 0.11           & 0.0127
  & 10       & 0.6           & 0.0684     \\
  $2^{12}$    & 0.25        & 173        & 0.030          & 0.0859
  & 10       & 0.6           & 0.325      \\
  $2^{13}$    & 0.125       & 665        & 0.0077         & 0.682
  & 10       & 0.6           & 1.08       \\
  $2^{14}$    & 0.0625      & 2674       & 0.0019         & 5.87
  & 10       & 0.6           & 3.07       \\
  $2^{15}$    & 0.03125     & 10514      & 0.00048        & 51.22
  & 10       & 0.6           & 9.75       \\
  \hline
  \end{tabular}

  \bigskip

  {\it Dirichlet BCs:}
  \smallskip

  \begin{tabular}{|c|c|c|c|c|c|c|c|}
  \hline
  Nodes $M$ & spacing $h$ & iterations ($L_2$) & step $\lambda$ ($L_2$)
  & CPUs ($L_2$)
  & iterations & step $\lambda$ & CPUs \\
  \hline
  \hline
  $2^{10}$    & 1.0         & 38       & 0.32            & 0.00586
  & 30           & 0.6            & 0.0146   \\
  $2^{11}$    & 0.5         & 115      & 0.11            & 0.0244
  & 33           & 0.6            & 0.0361    \\
  $2^{12}$    & 0.25        & 425      & 0.030           & 0.166
  & 52           & 0.6            & 0.159     \\
  $2^{13}$    & 0.125       & 1660     & 0.0077          & 1.32
  & 136          & 0.6            & 0.906     \\
  $2^{14}$    & 0.0625      & 6730     & 0.0019          & 11.64
  & 370          & 0.6            & 5.04     \\
  $2^{15}$    & 0.03125     & 26643    & 0.00048         & 105.33
  & 1029         & 0.6            & 29.69    \\
  \hline
  \end{tabular}

  \bigskip
  For small systems with large spacings the time taken by the
  solver negates the advantage of being able to use a much larger step
  $\lambda$ when using a Sobolev gradient.  However, as the system becomes
  larger and the spacing finer, the Sobolev gradient technique is more
  efficient.

  \bigskip

  {\bf Two dimensional model A}

  \bigskip

  Systems now have $M^2$ nodes.

  {\it Periodic BCs:}

  \smallskip

  \begin{tabular}{|c|c|c|c|c|c|c|c|}
  \hline
  $M$ & $h$ & iterations ($L_2$) & $\lambda$ ($L_2$) & CPUs ($L_2$)
  & iterations & $\lambda$ & CPUs \\
  \hline
  \hline
  $2^5$ & 1.00   & 27   & 0.19    & 0.005859
  & 10  & 0.6   & 0.0107 \\
  $2^6$ & 0.50   & 90   & 0.056   & 0.0576
  & 10  & 0.6   & 0.0693 \\
  $2^7$ & 0.25   & 332  & 0.015   & 0.939
  & 10  & 0.6   & 0.709  \\
  $2^8$ & 0.125  & 985  & 0.0038  & 14.58
  & 10  & 0.6   & 7.52      \\
  $2^9$ & 0.0625 & 3846 & 0.00097 & 301
  & 10  & 0.6  &  77.7       \\
  \hline
  \end{tabular}

  \bigskip

  {\it Dirichlet BCs:}

  \smallskip
  \begin{tabular}{|c|c|c|c|c|c|c|c|}
  \hline
  $M$ & $h$ & iterations ($L_2$) & $\lambda$ ($L_2$) & CPUs ($L_2$)
  & iterations & $\lambda$ & CPUs \\
  \hline
  \hline
  $2^5$ & 1.00   & 77    & 0.19    & 0.0127
  & 36    & 0.6   & 0.0263 \\
  $2^6$ & 0.50   & 263   & 0.056   & 0.15
  & 39    & 0.6   & 0.181  \\
  $2^7$ & 0.25   & 989   & 0.015   & 2.58
  & 83    & 0.6   & 2.46  \\
  $2^8$ & 0.125  & 3909  & 0.0038  & 53.09
  & 207   & 0.6   & 28.38  \\
  $2^9$ & 0.0625 & 15306 & 0.00097 & 1210.78
  & 640   & 0.6   & 387.78 \\
  \hline
  \end{tabular}

  \bigskip
  Again we note that the finer the spacing the less CPU time the Sobolev gradient
  technique uses in comparison to the usual steepest descent.
  For model A results in two dimensions the same step size $\lambda$ can
  be used for all spacings $h$ when minimizing in the
  appropriate Sobolev space.  The step size for minimization in $L_2$
  has to decrease as the spacing is refined, we note that it
  has to decrease much faster in two dimensions than in one.

  \bigskip
  \bigskip

  {\bf Three dimensional model A}

  \bigskip

  Systems now have $M^3$ nodes.

  {\it Periodic BCs:}

  \smallskip
  \begin{tabular}{|c|c|c|c|c|c|c|c|}
  \hline
  $M$ & $h$ & iterations ($L_2$) & $\lambda$ ($L_2$) & CPUs ($L_2$) &
  iterations & $\lambda$ & CPUs    \\
  \hline
  \hline
  $2^5$ & 1.00   & 36    & 0.14   & 0.303
  & 8   & 0.6  & 0.676     \\
  $2^6$ & 0.50   & 124   & 0.40   & 7.99
  & 8   & 0.6  & 7.55      \\
  $2^7$ & 0.25   & 494   & 0.010  & 429.16
  & 14  & 0.6  & 91.64     \\
  \hline
  \end{tabular}

  \bigskip

  {\it Dirichlet BCs:}

  \smallskip
  \begin{tabular}{|c|c|c|c|c|c|c|c|}
  \hline
  $M$ & $h$ & iterations ($L_2$)  & $\lambda$ ($L_2$) & CPUs ($L_2$) & iterations
  & $\lambda$ & CPUs    \\
  \hline
  \hline
  $2^5$ & 1.00   & 119   & 0.14    & 0.857
  & 41   & 0.6    & 2.32     \\
  $2^6$ & 0.50   & 417   & 0.040   & 27.57
  & 55   & 0.6   & 25.12     \\
  $2^7$ & 0.25   & 2115  & 0.010   & 1395.67
  & 171  & 0.6   & 591.31    \\
  \hline
  \end{tabular}

  \bigskip
  The three dimensional models also show that as the spacing becomes
  finer it is advantageous to use the Sobolev gradient technique.
    We also note from the preceding Tables that as the
  dimension of the problem increases the Sobolev gradient technique
  becomes more efficient.  In one dimension Sobolev gradients are
  more efficient for a spacing $h=0.25$, as compared to three dimensions
  where they are more efficient for spacing $h=0.5$.

  \section{\bf Conservation constraint }

  For Model A type systems the order parameter $u$ is not conserved.  A
  Cahn-Hilliard \cite{Cahn} or Model B system which would conserve the
  order parameter is given by

  $$
  u_t ~=~  \Gamma \nabla^2 \left[\frac{\delta J( u )}{\delta u}\right].
  $$

  Suppose  we wish to find the minima of some Model A type functional and
  we require conservation of the order parameter $u$ during the course of the
  simulation, without regard to the actual dynamics.
  We can use a second projection operator to enforce
  conservation rather than increase the order of our evolution equation by two.

  In order that $\int u du$ not change, we need to project our gradient onto the
  subspace of $L_2( \Omega )$ functions with integral zero.  This is achieved for
 a
  function
  $f$ by

  $$
  f ~\rightarrow~ f ~-~ \frac{ \int f }{ V }
  $$

  \noindent  We will use the term model A$^\prime$ for model A with this
  constraint as we do not solve for Model B dynamics.
  The order parameter $u$ is now taken to be a relative concentration of two
  fluids $A$ and $B$ with concentrations $\rho_A$, $\rho_B$, such that
  $\rho ~=~ \rho_A ~+~ \rho_B $ and $u ~=~ \frac{ \rho_A ~-~ \rho_B }{ \rho } $.

  %\noindent The free energy is given by

  We use the free energy

  $
  J ~=~ \int [ \frac{ \alpha }{ 4 } ( 1 ~-~ u^2 ) ~-~ T ~+~ \frac{T}{2} ( 1 ~+~ u
 )
  \log( 1/2 ~+~ u/2 ) ~+~ \frac{T}{2} ( 1 ~-~ u )
  \log( 1/2 ~-~ u/2 ) ~+~ $

  $ \ \ \ \ \ \ \ \ \ \  \ \ \ \ \ \ \ \ \
  \ \frac{ \kappa }{ 2 } | \nabla u |^2 ] \ \ \ \ \ \
   \ \ \ \ \ \ \ \ \ \ \ \ \ \ \ \ \ \ \ \ \ \ \ \
  \ \ \ \ \ \ \ \ \ \ \ \ \ \ \ \ \ \  \ \ \  \ \ \ \ \ \ \ (3)
  $

 \smallskip

  \noindent  This free energy contains the entropy of mixing.  Phase separation
  depends on the temperature $T$.  When $T$ is greater than the critical
  temperature $T_c = \alpha/2$ the two phases mix completely. When $T$
  is less than $T_c$ there will be domains of positive and negative
  $u$.  The lower $T$ is, the greater can be the possible maximum values of
  $|u|$ at equilibrium.  That is, phase separation between fluids
  $A$ and $B$ is more complete at lower $T$ values.

  %\section{\bf Results for binary systems }
  \smallskip

  The model B approach would result in an increase in the order of the
  derivatives of the evolution scheme by two.  Imposing conservation
  through a projection means that this can be avoided.  As a result,
  a Sobolev gradient approach for modeling systems with conservation
  constraints is even more efficient in comparison to the usual approach.
  The step size need not be reduced for finer spacings when using a Sobolev
  gradient scheme.  Minimization was performed on systems with random
  initial conditions and $\langle u \rangle = 0.05$, and $\alpha=2$, $T=0.8$,
 $\kappa=1.0$
  until the magnitude of the $L_2$ gradient was less than $10^{-5}$
  everywhere.  By comparing results in two and three dimensions we
  noticei from the Tables that the Sobolev gradient scheme is even more efficient
 in three
  dimensions than it is in two when compared to the traditional approach.

  %\vfill

  {\bf Two dimensional binary system with periodic BCs:}

  \bigskip
  \begin{tabular}{|c|c|c|c|c|c|c|c|}
  \hline
  $M$ & $h$ & iterations ($L_2$) & $\lambda$ ($L_2$) & CPUs ($L_2$)
  & iterations & $\lambda$ & CPUs \\
  \hline
  \hline
  $2^5$  &  1.00  & 680 000    & 0.027 & 50.34
  & 314  & 0.95 & 0.433  \\
  $2^6$  &  0.50  & 2 516 565  & 0.0018 & 740
  & 645  & 0.95 & 5.26    \\
  $2^7$  &  0.250 & 4 420 185  & 0.00012 & 5187
  & 1937 & 0.95 & 98.63   \\
  \hline
  \end{tabular}

  \bigskip

  {\bf Three dimensional binary system with periodic BCs:}

  \bigskip
  \begin{tabular}{|c|c|c|c|c|c|c|c|}
  \hline
  $M$ & $h$ & iterations & $\lambda$ & CPUs
  & iterations & $\lambda$ & CPUs \\
  \hline
  \hline
  $2^5$  &  1.00  & 418 515  & 0.012    & 6291
  & 323  & 0.95  & 33.23  \\
  $2^6$  &  0.50  & 594 233  & 0.00086  & 68523
  & 214  & 0.95  & 418    \\
  \hline
  \end{tabular}

  \smallskip

  These numerical experiments with model A$^\prime$ demonstrate that it
  is considerably more efficient to use a projection to enforce
  conservation of the order parameter if the final equilibrium configuration is
  all that is important.

  \section{\bf Surfactant systems }

  %\subsection{\bf Model }

  The addition of a surfactant to an oil-water system can be modeled by
  allowing $\kappa$ become negative \cite{Gompper} in the free energy (3).
  This favors the presence of interfaces between the two components
  of the mixture and thus mimics the action of surfactant in allowing
  the oil and water to ``mix" with the formation of bicontinuous domains
  separating the oil and water.
  We also add a curvature dependent term for a bending energy of the form
  $$
  E_{b} = \frac{ \gamma }{ 2 } ( \nabla^2 u )^2
  $$
  \noindent to the binary system free energy.  By changing $\gamma$ one can
  change the shape of domains from circular to oval.
  The surfactant model enables us to
  examine how the Sobolev gradient approach and the traditional schemes
  compare when the order of the derivatives increases.  The coefficient
  $\gamma$ was set to 1.0 and other parameters and initial conditions
  were as given in Section 5.

\bigskip

  {\bf Two dimensional surfactant system with periodic BCs:}

  \bigskip
  \begin{tabular}{|c|c|c|c|c|c|c|c|}
  \hline
  $M$ & $h$ & iterations ($L_2$) & $\lambda$ ($L_2$)  & CPUs ($L_2$)
  & iterations & $\lambda$ & CPUs \\
  \hline
  \hline
  $2^5$  &  1.00  &  4 853 277 & 0.0043     &  4 696
  & 43 234  & 0.5  & 336  \\
  $2^6$  &  0.50  & 27 103 876 & 0.000062   & 45 250
  &  4 798  & 0.5  & 449    \\
  $2^7$  &  0.250 & 96 649 780 & 0.00000096   & 97 327
  &  5 450  & 0.5  & 2038   \\
  \hline
  \end{tabular}

  \bigskip

  {\bf Three dimensional surfactant system with periodic BCs}

  It is clear that a model B minimization with sixth order
  derivatives will be much slower than using model A$^\prime$. We
  report results for the Sobolev gradient technique only.

  \bigskip
  \begin{tabular}{|c|c|c|c|c|c|c|c|}
  \hline
  $M$ & $h$ & iterations & $\lambda$   & CPUs  \\
  \hline
  \hline
  $2^5$  &  1.00  &  30 320     & 0.5    &  6  636 \\
  $2^6$  &  0.50  &  55 268     & 0.5    & 630 839 \\
  \hline
  \end{tabular}

  \section{\bf Summary and conclusions }

  We have presented minimization schemes for model A Ginzburg-Landau
  functionals  based on the Sobolev gradient technique \cite{SV,Adams}. The
 Sobolev gradient
  technique is computationally more efficient than the usual steepest descent
  method as the spacing of the numerical grid is made finer, the dimension of
  the problem is increased, the order of the derivatives in the functional is
  increased, or a conservation constraint is imposed.
  Our results indicate that Sobolev gradient techniques may offer distinct
  advantages in certain cases, particularly for problems involving functionals
  that contain spatial gradients such as strain based elasticity problems
  \cite{shenoy}, least
  square formulations of partial differential equations, and electrostatic
  problems that require solving the Poisson-Boltzmann equation.

  An interesting question is whether there exists
  an optimal metric with respect to which the
  Sobolev gradient works best.  It is an open research problem to try to
  find such an optimal metric, even though the optimal one would likely
  not make a large difference computationally in all cases.
  An example of where there is a great difference is in
  near-singular problems where a  weighted Sobolev gradient,
  weighted with the singularity in question, works vastly better
  \cite{Mahavier}. The likely fact that we cannot yet find an optimal
  metric may well be responsible for the nonlinear dependence of run time on
  number of grid points noted in this work.

  %%%%%%%%%%%%%%%%%%%%%%%%%%%%%%%%%%%%%%%%%%%%%%%%%%%%%%%%%%%%%%%%
  %%%%%%%%%%%%%%%%%%%%%%%%%%%%%%%%%%%%%%%%%%%%%%%%%%%%%%%%%%%%%%%%
  %%%%%%%%%%%%%%%%%%%%%%%%%%%%%%%%%%%%%%%%%%%%%%%%%%%%%%%%%%%%%%%%

  %%%% Authors begin text of article here %%%
  %%%% Authors begin text of article here %%%
  %%%% Authors begin text of article here %%%


\begin{thebibliography}{99}

  %% book
  \bibitem{SV}
  J.W. Neuberger,
  \textit{Sobolev Gradients and Differential Equations},
  Springer Lecture Notes in Mathematics 1670
  (Springer-Verlag, New York, 1997).

  % published journal article
  \bibitem{Hohenberg}
  P.C. Hohenberg, B.I. Halperin,
  Theory of dynamic critical phenomena ,
  \textit{Rev. Mod. Phys.}
  \textbf{49}, 435-479 (1977).

  % published journal article
  \bibitem{Richardson}
  W. B. Richardson,
  Steepest descent using smoothed gradients,
  \textit{Applied Mathematics and Computation}
  \textbf{112}, 241-254 (2000).

  % published journal article
  \bibitem{CFL}
  R. Courant, K. O. Friedrichs, H. Lewy,
  Uber die Partiellen Differenzengleichungen der Mathematisches Physik,
  \textit{Math. Ann. }
  \textbf{100}, 32-74 (1928).

  \bibitem{Adams}
  R.A. Adams, \textit{Sobolev Spaces}
  (Academic Press, New York, 1975).

  % published journal article
  \bibitem{Rogers}
  T.M. Rogers, K.R. Elder, R.C. Desai,
  Numerical study of the late stages of spinodal decomposition,
  \textit{Phys. Rev. B.}
  \textbf{37},
  (1988) 9638-9649.

  % published journal article
  \bibitem{Cahn}
  J.W. Cahn, J.E. Hilliard, Free Energy of a Nonuniform System. I.
 Interfacial Free Energy,
  \textit{J. Chem. Phys.}
  \textbf{28},
  (1958) 258.

  % published journal article
  \bibitem{Gompper}
  G. Gompper, M. Schick,
  Correlation between structural and interfacial properties of amphiphilic
      systems,
  \textit{Phys. Rev. Lett.}
  \textbf{65},
  (1990) 1116.
  % published journal article
  \bibitem{shenoy}
  S.R. Shenoy, T. Lookman, A. Saxena, A.R. Bishop, Martensitic
 Textures: Multiscale Consequences of elastic compatibility, Phys. Rev. B,
 v 60(18) pp. R12537-R12541 (1999).

  \bibitem{Mahavier}
  W.T. Mahavier, A Numerical Method Utilizing Weighted Sobolev Descent to
 Solve Singular Differential Equations,
 \textit{Nonlinear World} \textbf{4}, (1997) 4.

  %% conference procedings
  %\bibitem{Toro}
  %E. F. Toro,
  %Riemann-problem based techniques for computing reactive two-phase flows,
  %in
  %\textit{Proc. Third Intern. Confer. on Numerical Combustion, Antibes, France,
  %May 1989,}
  %edited by Dervieux and Larrouturrou,
  %Lect. Notes in Phys.
  %(Springer-Verlag, New York/Berlin, 1989),
  %Vol. 351, p. 472

  \end{thebibliography}
  \end{document}